\documentclass[nofootinbib]{revtex4}
\usepackage{graphicx}
\usepackage{latexsym}
\def\be{\begin{equation}}
\def\ee{\end{equation}}
\def\bea{\begin{eqnarray}}
\def\eea{\end{eqnarray}}

\def\ml{\mathcal{L}}

\begin{document}

\title{On dark energy models of single scalar field}
\author{Mingzhe Li$^{1,6}$}
\email{limz@nju.edu.cn}
\author{Taotao Qiu$^{2,3}$}
\email{xsjqiu@gmail.com}
\author{Yifu Cai$^{4}$}
\email{ycai21@asu.edu}
\author{Xinmin Zhang$^{5}$ }
\email{xmzhang@ihep.ac.cn} \affiliation{${}^1$ Department of
Physics, Nanjing University, Nanjing 210093, P.R. China}
\affiliation{${}^2$ Department of Physics and Center for Theoretical
Sciences, National Taiwan University, Taipei 10617, Taiwan}
\affiliation{${}^3$ Leung Center for Cosmology and Particle
Astrophysics National Taiwan University, Taipei 106, Taiwan}
\affiliation{${}^4$ Department of Physics, Arizona State University,
Tempe, AZ 85287, USA} \affiliation{${}^5$ Institute of High Energy
Physics, Chinese Academy of Sciences, P.O. Box 918-4, Beijing
100049, P.R. China} \affiliation{${}^6$ Joint Center for Particle,
Nuclear Physics and Cosmology, Nanjing University - Purple Mountain
Observatory, Nanjing 210093, P.R. China}

%\date{\today.}

\begin{abstract}
In this paper we revisit the dynamical dark energy model building based on single scalar field involving
higher derivative terms. By imposing a degenerate condition on the higher derivatives in curved spacetime, one can select
the models which are free from the ghost mode and the equation of state is able to cross
the cosmological constant boundary smoothly, dynamically violate the null energy condition. Generally the Lagrangian of this type of
dark energy models depends on the second derivatives linearly. It behaves like an imperfect fluid, thus its cosmological
perturbation theory needs to be
generalized. We also study such a model with explicit form of degenerate Lagrangian and show that its equation of state may
cross $-1$ without any instability.
\end{abstract}

\maketitle

\hskip 1.6cm PACS number(s): 98.80.Cq. \vskip 0.4cm

\section{Introduction}

The recent data from type Ia Supernovae (SNIa) and cosmic microwave
background (CMB) radiation and so on have provided strong evidences
for a spatially flat and accelerated expanding universe. In the
context of Friedmann-Robertson-Walker (FRW) cosmology with Einstein
gravity, this acceleration is attributed to the domination of a
component with negative pressure, called dark energy. So far, the
nature of dark energy remains a mystery. Theoretically, the simplest
candidate for such a component is a small positive cosmological
constant, but it suffers the difficulties associated with the fine
tuning and the coincidence problems. Therefore, many physicists are
attracted by the idea of dynamical dark energy models, such as
quintessence \cite{Ratra:1987rm,Wetterich:1987fm,Caldwell:1997ii},
phantom \cite{Caldwell:1999ew}, k-essence
\cite{ArmendarizPicon:2000dh}, quintom \cite{Feng:2004ad} and so on
(see Refs. \cite{Copeland:2006wr, Frieman:2008sn, Caldwell:2009ix,
Cai:2009zp, Li:2011sd} for recent reviews).

Although the recent fits to the data \cite{Li:2011dr}, in
combination of the 7-year WMAP \cite{Komatsu:2010fb}, the Sloan
Digital Sky Survey \cite{Reid:2009xm}, and the recently released
¡°Union2¡± SNIa data \cite{Riess:2011yx}, show remarkably the
consistence of the cosmological constant, it is worth noting that a
class of dynamical models with the equation-of-state (EoS) across
$-1$, dubbed as {\it quintom} which dynamically violates the null
energy condition (NEC), is mildly favored \cite{Feng:2004ad}.
However, it was noticed that consistent single field models
realizing the quintom scenario are difficult to be constructed. For
example, the EoS of quintessence is limited to be in the region
$-1\leq w \leq 1$, while for phantom $w$ is always smaller than
$-1$. It was proved that in models described by a single perfect
fluid or a single scalar field with a Lagrangian of k-essence form
\cite{ArmendarizPicon:1999rj}, the cosmological perturbations
encounter a divergence when the background EoS crosses $-1$
\cite{Feng:2004ad, Vikman:2004dc, Hu:2004kh, Caldwell:2005ai,
Zhao:2005vj}. This statement was explicitly proven in Ref.
\cite{Xia:2007km} as a ``No-Go" theorem for dynamical dark energy
models.

To realize a viable quintom scenario, one usually needs to add more degrees of freedom into the dark energy budget.
The simplest and also
the first quintom model was constructed by a combination of a
canonical scalar and a phantom scalar field \cite{Feng:2004ad}.
However theorists are still interested in pursuing single field quintom models.
The first single scalar field quintom model was realized by
introducing higher order derivative terms \cite{Li:2005fm}, also see
\cite{Zhang:2006ck} for generalization. It
was also obtained in the frame of nonlocal string theory
\cite{Aref'eva:2005fu} and decaying tachyonic branes
\cite{Cai:2007gs}. However, there is a quantum instability due to an
unbounded vacuum state in such type of models \cite{Carroll:2003st,
Cline:2003gs}. A possible approach to stable violations of NEC is the ghost condensation of Ref.
\cite{ArkaniHamed:2003uy}, in which the negative kinetic modes are
bounded via a spontaneous Lorentz symmetry breaking, although it
might allow for superluminal propagation of information in some
cases \cite{Dubovsky:2005xd}. Various theoretical realizations of
quintom scenarios and their implications for early universe physics
were reviewed in Ref. \cite{Cai:2009zp} (see also
\cite{Qiu:2010ux}).

Recently, a scalar field model which stably violates the NEC has
been studied extensively, which is the so-called Galileon
\cite{Nicolis:2008in}. It was viewed as a local infrared
modification of General Relativity, generalizing an effective field
description of the DGP model \cite{Dvali:2000hr}. The key feature of
these models is that they contain higher order derivative terms in
the action while the equation of motion remains second-order in
order to avoid the appearance of ghost modes, realizing the idea
pioneered by Horndeski thirty years ago \cite{Horndeski:1974}. Later
on, various phenomenological studies of this type of models were
performed, namely, see Refs. \cite{deRham:2010eu, Deffayet:2010qz,
Pujolas:2011he, Deffayet:2009wt, Deffayet:2011gz,
Gao:2011qe,Chow:2009fm, Silva:2009km, DeFelice:2010pv,
Kobayashi:2010cm, Hinterbichler:2010xn, Creminelli:2010ba}.
Motivated by the feature of the Galileon model, in this paper we revisit quintom
dark energy models containing higher derivative terms. We start from a general covariant Lagrangian of single
scalar field involving higher derivative terms and pursue how to keep the model free from extra degree of freedom.
We show that it is able to
eliminate the ghost mode by imposing a degenerate condition that
the Lagrangian only depends on the second derivative terms linearly. Based on the degenerate Lagrangian, we build
an explicit dark energy model and study its dynamics of its homogeneous background and its perturbations.
Our numerical calculations show that the EoS is able to cross the
cosmological constant boundary smoothly and the perturbation modes are well controlled when the crossing takes place.
We understand the reason of realizing the single field quintom without ghosts is that,
such a single scalar field model is no longer be
able to correspond to a perfect fluid. Thus, this model does not conflict with
the ``No-Go" theorem for quintom dark energy model building as proposed in
\cite{Xia:2007km}.

This paper is organized as follows. In Section II, we simply review the difficulty of constructing
a single field dark energy model which gives rise to quintom scenario. In Section III, we start
with a single field action involving higher derivative terms, and present the general analysis.
We also discuss under what condition the higher derivative terms do not bring a ghost mode to the
effective field description in Section IV. Section V is devoted to the study of the background and
perturbation dynamics of single field dark energy model with degenerate higher derivatives in the frame
of flat FRW universe. Specifically, we present an explicit model of quintom dark energy with degenerate
higher derivatives. We perform numerical computation to illustrate such a model can realize the EoS across $-1$ smoothly.
Section VI is the summary.

\section{The difficulty of single field dark energy models with EoS crossing $-1$}

We begin by briefly reviewing the difficulty of constructing
dynamical dark energy models with the EoS across the cosmological
constant boundary. As demonstrated by several groups
\cite{Feng:2004ad, Vikman:2004dc, Hu:2004kh, Caldwell:2005ai,
Zhao:2005vj, Xia:2007km}, the EoS of dark energy based on single
perfect fluid or single k-essence scalar field (quintessence and
phantom are special cases of k-essence) cannot cross $-1$ remaining finite perturbations.
The key point to the proof of this
no-go theorem is that in both scenarios the pressure perturbation
has a gauge invariant relation with energy and momentum density
perturbations, \be\label{start}
 \delta p= c_s^2 \delta \rho+3H (c_s^2-\frac{\dot p}{\dot \rho})\frac{V}{k^2}~,
\ee where the dot represents the derivative with respect to time, and
$H=\dot a/a$ is the Hubble rate with $a$ the scale factor. The sound
speed square $c_s^2$ is defined as the ratio of $\delta p/\delta
\rho$ at the frame comoving with the dark energy, it should be
positive definite to guarantee the Jeans stability of
perturbations at small scales. The momentum density perturbation is
defined as $V=ik^i T^0_i$ in Fourier space, here $T^0_i$ is the $0-i$ component of the energy momentum tensor.
For perfect fluid the
pressure is a function of the energy density only, $p=p(\rho)$, and $c_s^2=\dot
p/\dot\rho$, hence $\delta p=(\dot p/\dot\rho)\delta\rho$. For
k-essence field the sound speed square is generally different from
$\dot p/ \dot\rho$, however, from Eq. (\ref{start}) we can see that
in both cases the pressure perturbation diverges at the crossing
point $\dot\rho=0$ with finite $V$ and density perturbation. As we learn from the gravitational field equation,
a divergent pressure perturbation will lead to arbitrarily large metric perturbations.
This instability can also be seen from the
equation of motion or the action for the perturbations.
A generic form of Lagrangian for a k-essence field
is only a function of the scalar $\phi$ and its first derivatives
$X=\frac{1}{2}\nabla_{\mu}\phi\nabla^{\mu}\phi$, of which the action
is expressed as, \be S=\int d^4x\sqrt{g} \ml (\phi,~X)~, \ee where
$g$ is the negative determinant of the metric tensor $g_{\mu\nu}$.
Its energy momentum tensor has the same form as that of perfect
fluid, \be T^{\mu\nu}=-pg^{\mu\nu}+(\rho+p)u^{\mu}u^{\nu}~, \ee
where \be
p=\ml~,~~\rho=-p+2Xp_X~,~~u^{\mu}=\frac{\nabla^{\mu}\phi}{\sqrt{2X}}~.
\ee We have used the subscript $X$ to indicate the partial
derivative with respect to $X$. The evolution of the k-essence
perturbation $\pi=\phi(\vec{x},~t)-\phi(t)$ is governed by the
equation which may be obtained via variational principle from the
following action of second order, \be\label{kessence}
S^{(2)}={1\over 2} \int d^3x dt a^3
[\rho_X\dot\pi^2-\frac{p_X}{a^2}\partial_i\pi\partial_i\pi+p_{\phi\phi}\pi^2-
\frac{d(a^3\dot\phi p_{X\phi})/dt}{a^3}\pi^2]~. \ee We have
neglected the metric perturbations for simplicity. We can see from
the action that the sound speed square is $c_s^2 =p_X/\rho_X$.
However, from the relation $\rho+p=2Xp_X$ we know
that $p_X$ vanishes at the point of crossing and changes the sign
after the crossing. To guarantee $c_s^2>0$, $\rho_X$ should also
vanish at the crossing point and change the sign afterwards, just as
$p_X$. The vanishment of $\rho_X$ will make the equation of perturbation singular and the
amplitude of the perturbation arbitrarily large around the crossing
point.

Such kind of instability is classical. Another difficulty emerges when the quantum effects are considered.
At the phantom phase $w<-1$, $p_X$ is negative
and $c_s^2>0$ requires $\rho_X<0$. The action (\ref{kessence})
showed that the kinetic term of $\pi$ has a wrong sign. This means
$\pi$ in this phase is a ghost which brings the problem of vacuum
instability due to the existence of negative energy states or of
violating unitarity by negative norm states.

\section{Single scalar dark energy model with higher derivatives}

To avoid the problem of singular perturbation possessed by single
k-essence field, earlier quintom model buildings
introduced multi-degree of freedom explicitly. For example in Ref.\cite{Feng:2004ad}, the quintom model is constructed by a
quintessence field and a phantom field. Though the total EoS of
these two fields crosses $-1$ during the evolution, each component
does not cross this boundary and has regular perturbation. Besides
the multi-fluid or multi-field models, it is still interesting and
important to pursue quintom models with single degree of freedom. To
this end, some extensions beyond the perfect fluid and k-essence
field are proposed in the literature including the higher derivative
field theory \cite{Li:2005fm}, the non-minimal coupling to the
gravity \cite{Boisseau:2000pr}, the constrained scalar field
which violates Lorentz invariance locally \cite{Lim:2010yk},
%the fluid with shear viscosity
%\cite{}
and so on. In this paper we only consider the first extension.

For a scalar field with higher (but finite) derivatives, its
Lagrangian generally has the form, \be
 \ml = \ml (\phi, \phi_{\mu_1}, \phi_{\mu_1\mu_2}, ..., \phi_{\mu_1 ...\mu_N})~,
\ee where $\phi_{\mu_1}\equiv
\nabla_{\mu_1}\phi,~\phi_{\mu_1\mu_2}\equiv
\nabla_{\mu_2}\nabla_{\mu_1}\phi$ and so on are the covariant
derivatives of $\phi$ and $N\geq 2$. The equation of motion from
this Lagrangian is \be
 \frac{\partial \ml}{\partial \phi} + \sum_{n=1}^N (-1)^n\nabla_{\mu_1} ... \nabla_{\mu_n}(\frac{\partial \ml}{\partial \phi_{\mu_1...\mu_n}}) = 0~.
\ee Generally this is a $2N$th order derivative equation, the whole
system contains $N$ degrees of freedom and some of them are ghosts.
In order to keep the discussions simple and without loss of general
properties of higher derivative field theories, we
only consider the case $N=2$ in curved spacetime, the Lagrangian is
a scalar function of $\phi,~\phi_{\mu}$ and $\phi_{\mu\nu}$. The
equation of motion is \be
 \frac{\partial \ml}{\partial \phi} - \nabla_{\mu}(\frac{\partial\ml}{\partial \phi_{\mu}}) +\nabla_{\mu}\nabla_{\nu}
 (\frac{\partial\ml}{\partial \phi_{\mu\nu}}) = 0~.
\ee Expanding this equation and considering the symmetry
$\phi_{\mu\nu}=\phi_{\nu\mu}$, we have the following equation,
\bea\label{expansion}
 & &\frac{\partial\ml}{\partial\phi} - \frac{\partial^2\ml}{\partial\phi\partial\phi_{\mu}}\phi_{\mu} + (\frac{\partial^2\ml}
 {\partial\phi\partial \phi_{\mu\nu}} - \frac{\partial^2\ml}{\partial\phi_{\nu}\partial \phi_{\mu}})\phi_{\nu\mu} +
 \frac{\partial^3\ml}{\partial\phi\partial\phi\partial\phi_{\mu\nu}}\phi_{\mu}\phi_{\nu} + 2\frac{\partial^3\ml}
 {\partial\phi\partial\phi_{\rho}\partial\phi_{\mu\nu}}\phi_{\rho\mu}\phi_{\nu} + \frac{\partial^3\ml}
 {\partial\phi_{\rho}\partial\phi_{\sigma} \partial\phi_{\mu\nu}}\phi_{\sigma\mu}\phi_{\rho\nu} + \nonumber\\
 & &\frac{\partial^2\ml}{\partial\phi_{\rho}\partial\phi_{\mu\nu}}(\phi_{\nu\rho\mu} -\phi_{\nu\mu\rho}) +
 2\frac{\partial^3\ml}{\partial\phi\partial\phi_{\rho\sigma} \partial\phi_{\mu\nu}} \phi_{\rho\sigma\mu}\phi_{\nu}
 + 2\frac{\partial^3\ml}{\partial\phi_{\alpha} \partial\phi_{\rho\sigma} \partial \phi_{\mu\nu}}\phi_{\rho\sigma\mu}\phi_{\alpha\nu} +
 \frac{\partial^3\ml}{\partial\phi_{\alpha\beta}\partial\phi_{\rho\sigma}\partial \phi_{\mu\nu}}\phi_{\alpha\beta\mu}\phi_{\rho\sigma\nu} +\nonumber\\
 & &\frac{\partial^2\ml}{\partial\phi_{\rho\sigma} \partial \phi_{\mu\nu}}\phi_{\rho\sigma\nu\mu} = 0~.
\eea If the matrix
$\frac{\partial^2\ml}{\partial\phi_{\rho\sigma}\partial
\phi_{\mu\nu}}$ is non-degenerate, this is a fourth order
differential equation. To solve this equation we need to impose the
values of $\phi,~\phi_{\mu},~\phi_{\mu\nu},~\phi_{\mu\nu\rho}$ at
the initial surface. This means such a model essentially possesses
two dynamical components.

The first dark energy model of single scalar field with higher derivatives
was proposed in \cite{Li:2005fm} where a simple example is taken to
illustrate the feature of crossing the boundary of $-1$. The
sample model has the Lagrangian of the type, \be\label{lagrangian}
 \ml=AX+B(\Box\phi)^2-V(\phi)~,
\ee where $\Box=\nabla_{\mu}\nabla^{\mu}$, $V(\phi)$ is the
potential and $A,~B$ are constants. In Ref. \cite{Li:2005fm}, it was
shown explicitly that this model is equivalent to the double-field
model with lower derivatives, but one field has negative kinetic
term. This can be shown as follows. Taking an integration by parts
the Lagrangian (\ref{lagrangian}) can be rewritten as a function
which does not depend on $X$. So it belongs to the more general
class in which the Lagrangian is an arbitrary scalar function of
$\phi$ and $\Box\phi$, \be\label{lagrangian1}
 \ml=\ml (\phi,~\Box\phi)~.
\ee Now we use this Lagrangian to illustrate how is the
non-degenerate higher derivative field model equivalent to
multi-field model with lower derivatives. Non-degeneracy means
$\ml_{BB}\neq 0$, where the subscript $B$ represents the derivative
with respect to $\Box\phi$. We first introduce an auxiliary field
defined as \be\label{aux} \varphi\equiv \ml_{B}~. \ee Due to the
non-degeneracy, we may convert the above equation to get $\Box\phi$ as a
function of $\phi$ and $\varphi$. Then change the independent
variables $(\Box\phi,~\phi)$ to $(\phi,~\varphi)$, we accordingly
have the Legendre transform of the Lagrangian in Eq.
(\ref{lagrangian1}), \be U(\phi,~\varphi)=\phi\Box \phi
(\phi,~\varphi)-\ml~, \ee it may be considered as a potential of
$\phi$ and $\varphi$. After finding the potential $U$, the higher
derivative term in the Lagrangian (\ref{lagrangian1}) can be removed
with the price of introducing extra field, \be \ml=\varphi \Box\phi
-U( \phi,~\varphi)~, \ee which is equivalent to the following
Lagrangian \be \ml'=-\nabla_{\mu}\varphi\nabla^{\mu}\phi-U(
\phi,~\varphi)~. \ee This equivalent Lagrangian has a more clear
form of double fields, \be\label{lagrangian2}
\ml'=\frac{1}{2}\nabla_{\mu}\phi_1\nabla^{\mu}\phi_1-
\frac{1}{2}\nabla_{\mu}\phi_2\nabla^{\mu}\phi_2- U(\phi_1,~\phi_2)~,
\ee through the field redefinitions \be\label{redefinition}
\phi_1=\frac{1}{\sqrt{2}}(\phi-\varphi)~,~~~~\phi_2=\frac{1}{\sqrt{2}}(\phi+\varphi)~.
\ee The mode $\phi_2$ is a ghost which violates the null energy
condition because its kinetic term has a wrong sign.
 This explains why the model (\ref{lagrangian1}) may cross the cosmological constant boundary without divergent perturbation.
 However at the quantum level, the non-degenerate higher derivative model is plagued by the existence of ghost mode.

\section{Degenerate higher derivative model}

In the degenerate higher derivative model, the matrix
$\frac{\partial^2\ml}{\partial\phi_{\rho\sigma}\partial
\phi_{\mu\nu}}$ in Eq. (\ref{expansion}) is identically zero.
Correspondingly, all the third and fourth order derivative terms in
the equation of motion (\ref{expansion}) disappear, i.e.,
\bea\label{expansion2} &
&\frac{\partial\ml}{\partial\phi}-\frac{\partial^2\ml}{\partial\phi\partial\phi_{\mu}}\phi_{\mu}+
(\frac{\partial^2\ml}{\partial\phi\partial
\phi_{\mu\nu}}-\frac{\partial^2\ml}{\partial\phi_{\nu}\partial
\phi_{\mu}})\phi_{\nu\mu}
+\frac{\partial^3\ml}{\partial\phi\partial\phi\partial\phi_{\mu\nu}}\phi_{\mu}\phi_{\nu}
+2\frac{\partial^3\ml}{\partial\phi\partial\phi_{\rho}\partial\phi_{\mu\nu}}\phi_{\rho\mu}\phi_{\nu}+\frac{\partial^3\ml}
{\partial\phi_{\rho}\partial\phi_{\sigma}
\partial\phi_{\mu\nu}}\phi_{\sigma\mu}\phi_{\rho\nu}+\nonumber\\
& &\frac{\partial^2\ml}{\partial\phi_{\rho}\partial \phi_{\mu\nu}}
\phi_{\sigma}R^{\sigma}_{~~\nu\mu\rho}=0~, \eea where we have used
the commutation of the covariant derivatives
$\phi_{\nu\rho\mu}-\phi_{\nu\mu\rho}=\phi_{\sigma}R^{\sigma}_{~~\nu\mu\rho}$
with $R^{\sigma}_{~~\nu\mu\rho}$ the Riemann tensor. The equation of
motion remains to be a second order differential equation, but there
is a curvature-field coupling term appeared in it even though we
only consider the minimal coupling to the gravity in the Lagrangian.
This means the degenerate model has no extra degree of freedom.
\footnote{ In the flat spacetime it is not necessary to require
$\frac{\partial^2\ml}{\partial\phi_{\rho\sigma}\partial
\phi_{\mu\nu}}=0$ to keep the equation of motion at the second
order, for example if
$\frac{\partial^2\ml}{\partial\phi_{\rho\sigma}\partial
\phi_{\mu\nu}}=C(\eta^{\mu
\nu}\eta^{\rho\sigma}-\eta^{\mu\rho}\eta^{\nu\sigma})$ with constant
$C$ in the Minkowski space, all the third and fourth order derivative
terms vanish in the equation of motion because
$\phi_{\mu\nu\rho\sigma}$ is totally symmetric under the interchanges
of the indices. See Ref. \cite{Deffayet:2011gz} for some
more discussions about the case in flat spacetime. But in the curved spacetime,
$\frac{\partial^2\ml}{\partial\phi_{\rho\sigma}\partial
\phi_{\mu\nu}}=0$ is the unique way to discriminate higher
derivative terms. This has be shown in Ref. \cite{Horndeski:1974}.}

Zero matrix $\frac{\partial^2\ml}{\partial\phi_{\rho\sigma}\partial
\phi_{\mu\nu}}$ implies the Lagrangian only depends on the second
derivative terms linearly. With Lorentz invariance, $\ml$ can only
be \be\label{FG}
\ml=K(\phi,~X)+G(\phi,~X)\Box\phi+F(\phi,~X)\nabla^{\mu}\phi\nabla^{\nu}\phi\nabla_{\mu}\nabla_{\nu}\phi~.
\ee
The box term at the right hand side with $G=X$ was considered in the context of Galileon theory
\cite{Nicolis:2008in} and its generalization was studied in \cite{Deffayet:2010qz}, named as $KGB$ model
and in inflation model building \cite{Kobayashi:2010cm}, named as $G$-inflation.
The box term is
equivalent to
$-2XG_{\phi}-G_X\nabla^{\mu}\phi\nabla^{\nu}\phi\nabla_{\mu}\nabla_{\nu}\phi$
after integration by parts and dropping a surface term. By
redefinitions of $K(\phi,~X)$ and $F(\phi~,X)$, the degenerate
Lagrangian may be generally written as \be\label{kg}
\ml=K(\phi,~X)+F(\phi,~X)\nabla^{\mu}\phi\nabla^{\nu}\phi\nabla_{\mu}\nabla_{\nu}\phi=
K+F\nabla_{\mu}X\nabla^{\mu}\phi~. \ee
Now we will investigate whether the dark energy model from this Lagrangian can stably
cross the boundary of cosmological constant
by studying its background evolution and properties of perturbations.

With the notations (\ref{kg}) the equation of motion
(\ref{expansion2}) becomes \bea\label{eom4} &
&K_{\phi}-2XK_{X\phi}-K_X\Box\phi-K_{XX}\nabla_{\mu}X\nabla^{\mu}\phi
+F[(\Box\phi)^2-\nabla_{\mu}\nabla_{\nu}\phi\nabla^{\mu}\nabla^{\nu}\phi+R_{\mu\nu}\nabla^{\mu}\phi\nabla^{\nu}\phi]\nonumber\\
&+&2F_{\phi}(\nabla_{\mu}X\nabla^{\mu}\phi+2X\Box\phi)+2XF_{\phi
X}\nabla_{\mu}X\nabla^{\mu}\phi+4X^2F_{\phi\phi}
+F_X(\Box\phi\nabla_{\mu}X\nabla^{\mu}\phi-\nabla_{\mu}X\nabla^{\mu}X)=0~.
\eea The energy momentum tensor which sources the gravitational
field is obtained through the variation of the action with respect
to the metric tensor, \be T^{\mu\nu}=-\frac{2}{\sqrt g}\frac{\delta
S}{\delta g_{\mu\nu}}~, \ee for the degenerate model it is \be
T^{\mu\nu}=-(K+F\nabla_{\rho}X\nabla^{\rho}\phi)g^{\mu\nu}+
(K_X-2XF_{\phi}-F\Box\phi)\nabla^{\mu}\phi\nabla^{\nu}\phi
+F(\nabla^{\mu}X\nabla^{\nu}\phi+\nabla^{\nu}X\nabla^{\mu}\phi)~.
\ee We can read off the pressure and energy density from this energy
momentum tensor up to linear order of perturbations around the homogeneous background, \bea\label{pressure}
p &=& K+F\nabla_{\mu}X\nabla^{\mu}\phi~,\\
\label{density}\rho
&=&-K+F\nabla_{\mu}X\nabla^{\mu}\phi+2X(K_X-2XF_{\phi}-F\Box\phi)~.
\eea
In terms of the fluid variables $p$ and $\rho$, the energy momentum tensor may be rewritten as
\be\label{emt2} T^{\mu\nu}=-pg^{\mu\nu}+(\rho+p)u^{\mu}u^{\nu}
+(2X)^{3/2}F(a^{\mu}u^{\nu}+a^{\nu}u^{\mu})~, \ee where by analogy with k-inflation \cite{Garriga:1999vw} or k-essence \cite{ArmendarizPicon:2000dh}
we have defined the four
velocity $u^{\mu} = \nabla^{\mu} \phi/\sqrt{2X}$ which is
normalized as $u^{\mu}u_{\mu} = 1$, $a^{\mu} \equiv
u^{\rho}\nabla_{\rho}u^{\mu}$ is the four acceleration which is
orthogonal to the velocity, i.e., $a_{\mu}u^{\mu}=0$. We have also
used the relation \cite{Pujolas:2011he}: \be \nabla^{\mu}X=2X
a^{\mu}+u^{\rho}\nabla_{\rho}X u^{\mu}~. \ee This energy momentum
tensor (\ref{emt2}) does not have the form of perfect fluid due to
the last two terms depending on the four acceleration. In the language of
relativistic imperfect fluid, $a^{\mu}$ can be identified as heat flow.
Further, one can prove straitforwardly that $a^{\mu}$ is space like
and its zero-th component vanishes at both background and linear
levels, its spatial components $a_i$ should be first order variables. So if we redefine
the four velocities as \be
\tilde{u}^{\mu}=u^{\mu}+\frac{(2X)^{3/2}F}{\rho+p}a^{\mu}~, \ee the
energy-momentum tensor should be \be
T^{\mu\nu}=-pg^{\mu\nu}+(\rho+p)\tilde{u}^{\mu}\tilde{u}^{\nu}
-\frac{(2X)^{3}F^2}{\rho+p}a^{\mu}a^{\nu}\simeq
-pg^{\mu\nu}+(\rho+p)\tilde{u}^{\mu}\tilde{u}^{\nu}~. \ee In the
last step, we have considered the fact that the product
$a^{\mu}a^{\nu}$ is a second order variable and can be neglected if
we restrict our studies on the background evolution and the linear perturbation theory.
This equation means the energy momentum tensor of the degenerate model has apparently
the same form of perfect fluid if we neglect perturbations of higher order. But it is essentially
different from perfect fluid, especially the relation
(\ref{start}) between the pressure and
density perturbations is lost in this model. This is the very reason why the degenerate higher derivative model is possible to
realize the quintom idea and avoid the problems possessed by single k-essence field.

\section{Cosmology with degenerate higher derivative dark energy model}

In this section,
we consider in more detail the dynamics of the dark energy model (\ref{kg}). The background universe
is a spatially flat FRW spacetime, in which the metric
is given by \be
 ds^2=dt^2-a^2(t)\delta_{ij}dx^{i}dx^{j}~.
\ee
At the background level the field $\phi$ is homogeneous, so we
soon have \bea\label{rhop}
 &&\rho=-K+2X(K_X-2XF_{\phi}-3H\dot\phi F)\nonumber\\
 &&p=K+2X\ddot \phi F\nonumber\\
 &&\rho+p=2X(K_X-2XF_{\phi}-3H\dot\phi F+\ddot\phi F)~,
\eea where we have considered
$X=\dot\phi^2/2~,\nabla_{\mu}X\nabla^{\mu}\phi=2X\ddot\phi$ and $\Box\phi=\ddot\phi+3H\dot\phi$. The energy conservation law in the
expanding universe $\dot\rho+3H (\rho+p)=0$ is identical to the
equation of motion (\ref{eom4}). For the velocity $u^{\mu}$ only the
time component is non-zero, $u^0=1,~u^i=0$.
The key point for this model to cross the cosmological constant
boundary is that $K_X-2XF_{\phi}-3H\dot\phi F+\ddot\phi F$ could evolve from
the positive region to the negative region or vise versa providing
the energy density always positive. In addition, we have to check whether the perturbations are stable.

Similar to the analysis in single k-essence model, the perturbation of the scalar
field is a small deviation from the homogenous background \be
 \phi(\vec{x},~t)=\phi(t)+\pi(\vec{x},~t)~.
\ee For complete consideration we should also include the
perturbations of spacetime. However, for dark energy, it is believed to be subdominant in the universe for most time and had
tiny contribution to the curvature. So for studying the dark energy perturbations, it is safely to neglect the metric
perturbations which are sourced mainly by other matter. This approximation will greatly simplify the analysis.
There are two ways to get the linear perturbation equations. One is
to directly perturb the equation of motion (\ref{eom4}) around the
background evolution. Another one is to adapt the variational
principle from the action which is second order of perturbations as we have shown in Sec. II for k-essence field.
Here we will use the second way to discuss the perturbation of the
scalar field. For this purpose we firstly expand the action
\be\label{general}
 S=\int d^4x\sqrt{g}(K+F\nabla_{\mu}X\nabla^{\mu}\phi)
\ee to the second order of the perturbations. After straightforward
but tedious calculations, we obtain the desired second order action
for $\pi$, \bea
 S^{(2)}(\pi) &=& {1\over 2}\int d^3x dt a^3(A\dot\pi^2 -\frac{B}{a^2}\partial_i\pi\partial_i\pi+C\pi^2)~,
\eea with \bea\label{a}
 A &=& K_X+2XK_{XX}-6H\dot\phi (F+XF_X)-8XF_{\phi}-4X^2F_{X\phi}~,\nonumber\\
 B &=& K_X-2(\ddot\phi+2H\dot\phi)F-4XF_{\phi}-2X\ddot\phi F_X~,\nonumber\\
 C&=& \frac{d}{dt}(6HXF_{\phi}+2X\dot\phi F_{\phi\phi}-\dot\phi K_{X\phi})+3H(6HXF_{\phi}+2X\dot\phi F_{\phi\phi}-\dot\phi K_{X\phi})
 +2X\ddot\phi F_{\phi\phi}+K_{\phi\phi}~.
\eea We can see from the action that the sound speed square is
defined as \be\label{cs2}
 c_s^2 = \frac{B}{A} = \frac{K_X-2(\ddot\phi+2H\dot\phi)F-4XF_{\phi}-2X\ddot\phi F_X}{K_X+2XK_{XX}-6H\dot\phi (F+XF_X)-8XF_{\phi}-4X^2F_{X\phi}}~.
\ee

If the universe is dominated by the scalar field, as discussed in the $KGB$ model \cite{Deffayet:2010qz} or the
$G$-inflation model \cite{Kobayashi:2010cm}, a full treatment of the gravity-$\phi$ coupled
system based on the (Arnowitt-Deser-Misner) ADM method will give a slightly different sound speed squared,
 \be c_s'^2 =
\frac{K_X-2(\ddot\phi+2H\dot\phi)F-4XF_{\phi}-2X\ddot\phi F_X-2X^2
F^2/M_p^2} {K_X+2XK_{XX}-6H\dot\phi
(F+XF_X)-8XF_{\phi}-4X^2F_{X\phi}+6X^2F^2/M_p^2}~, \ee
where both
numerator and denominator are modified by terms suppressed by the
Planck mass $M_p^2=1/8\pi G$. This sound speed depends on the gravity theory, here the gravity theory is
Einstein's general relativity. For dark energy studied in this paper we will consider the sound
speed in Eq. (\ref{cs2}) to express the propagating velocity
of the perturbations.

The classical stability requires $c_s^2>0$. Furthermore, the absence
of ghost mode corresponds to \be
 A = K_X+2XK_{XX}-6H\dot\phi (F+XF_X)-8XF_{\phi}-4X^2F_{X\phi}>0~.
\ee Compared to Eq. (\ref{rhop}) one can find that neither $A$ nor
$B$ is proportional to $\rho+p$, hence when the dark energy crosses
the cosmological constant boundary, $\rho+p=0$, both coefficients $A$
and $B$ are not vanished in general. The equation of motion would be
regular at the crossing point. This is different from the case of
k-essence model discussed in Section II. So in this single field
model, for particular choices of the functions $K$ and $F$ and corresponding model parameters,
it is possible to find solutions in which the equation of
state of dark energy evolves across $-1$ but both $c_s^2$ and $A$ remain finite and
positive as we will show explicitly below.
Within these solutions the fluctuation of the scalar field has the right kinetic term to circumvent
the pathology of ghost, even though
its background part has $w<-1$ violating the null energy
condition.

In order to illustrate the realization of quintom scenario
explicitly, we study an explicit form of the
degenerate dark energy model. The Lagrangian we are considering is
simple: \be\label{conformal} \ml =
-X-c_1\nabla_{\mu}X\nabla^{\mu}\phi+c_2X\phi^2~, \ee where $c_{1,2}$
are constants. Compared with the notations in Eq. (\ref{kg}), one
may find that $K=-X+c_2X\phi^2$ and $F=-c_1$. Note that the third
term can also be viewed as an ``effective mass term". From Eqs.
(\ref{pressure}) and (\ref{density}), we get the pressure and energy
density of this model respectively as: \bea\label{densityp}
\rho&=&(c_2\phi^2-1+6c_1H\dot\phi)X~,\\
p&=&(c_2\phi^2-1-2c_1\ddot\phi)X~.\eea At the point where the
equation of state crossing $-1$, we have $\rho+p=0$, i.e., \be
 c_1(\ddot\phi-3H\dot \phi)=c_2\phi^2-1~.
\ee In addition, we have the equation of motion for the scalar
field, which comes from Eq. (\ref{eom4}), \be
 (1 - 6c_1H\dot\phi - c_2\phi^2) \ddot\phi + 3H(1-c_2\phi^2)\dot\phi - (c_2\phi + 9c_1H^2 + 3c_1\dot{H}) \dot\phi^2 = 0~.
\ee

\begin{figure}
\begin{center}
\includegraphics[scale=0.7]{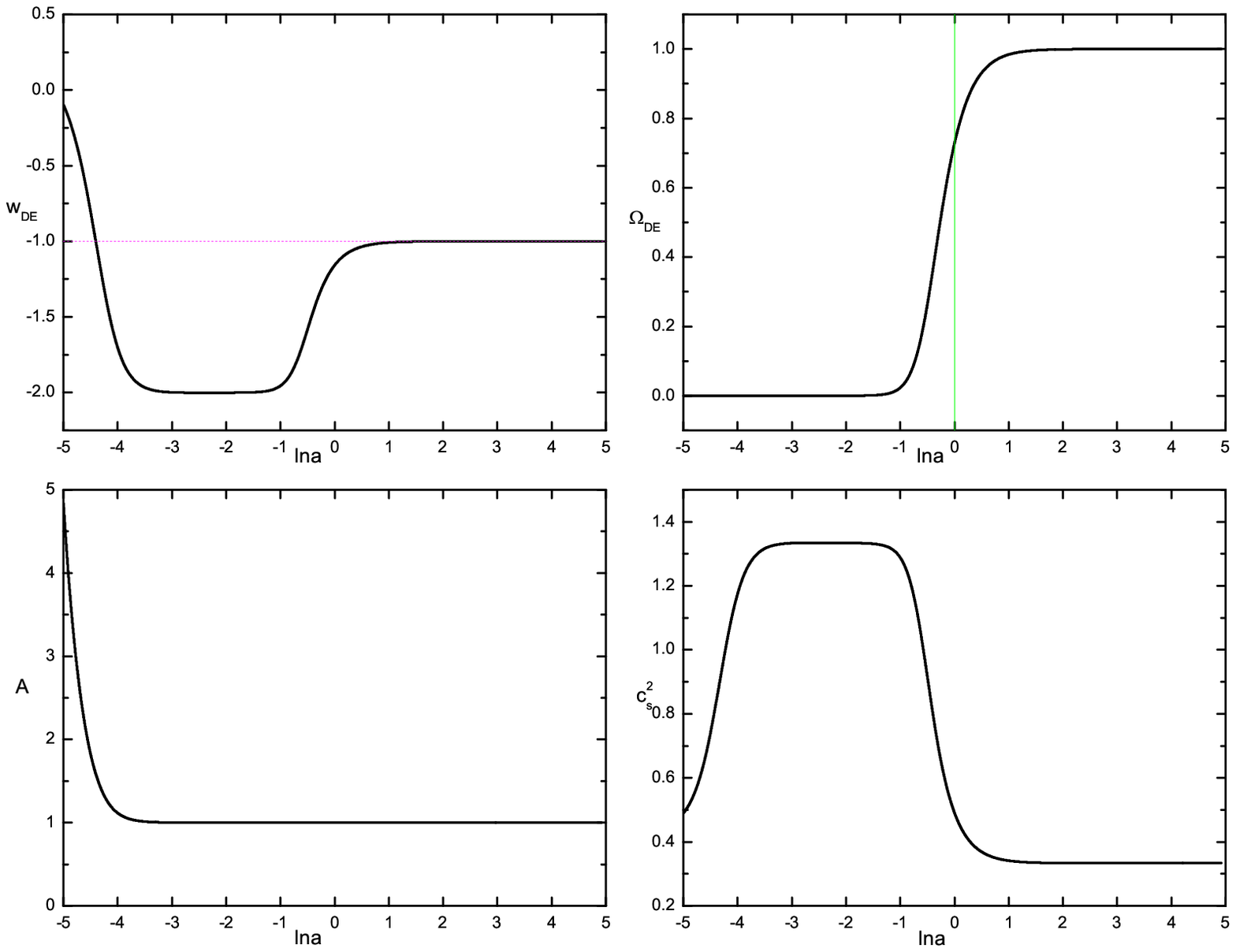}
\caption{The equation of state $w_{DE}$, the energy density ratio
$\Omega_{DE}$, parameter $A$ and sound speed squared $c_s^2$ of dark
energy of the model (\ref{conformal}). The parameters are chosen to be
$c_1 = 3.0 \times 10^{122}$, $c_2 = 0$ and the initial conditions
are chosen in the far past when the universe was matter dominated.
The equation of sate crosses $-1$ recently, and approach to $-1$ in
the future. Both $A$ and $c_s^2$ are positive during the evolution.
The universe will be dark energy dominant in the future.} \label{lmzrhoandw}
\end{center}
\end{figure}

\begin{figure}
\begin{center}
\includegraphics[scale=0.7]{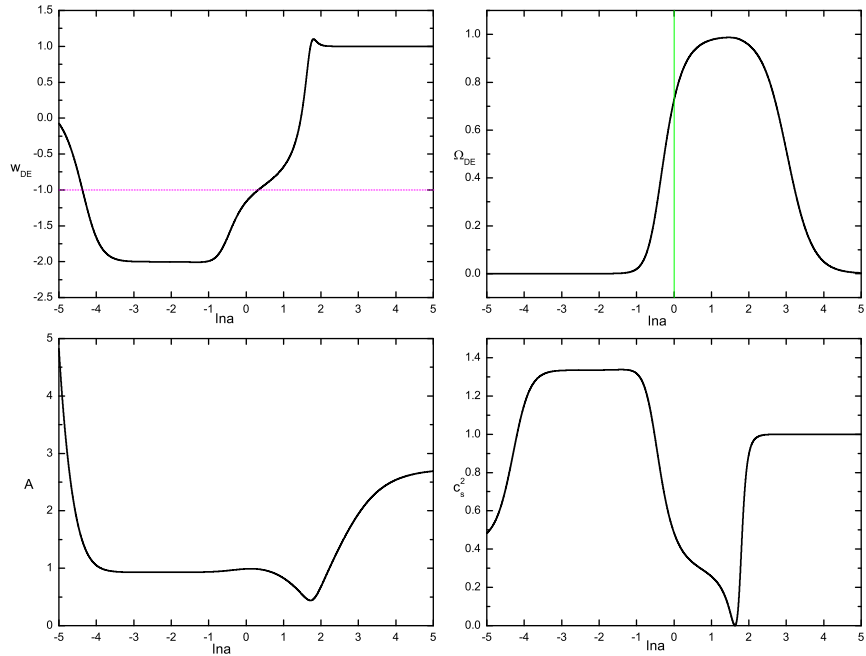}
\caption{The equation of state $w_{DE}$, the energy density ratio
$\Omega_{DE}$, parameter $A$ and sound speed squared $c_s^2$ of dark
energy of the model (\ref{conformal}). The parameters are chosen to be
$c_1 = 3.0 \times 10^{122}$, $c_2 = 0.05$ and the initial conditions
are chosen in the far past when the universe was matter dominated.
The equation of sate crosses $-1$ recently, and becomes larger than
$0$ in the future. Both $A$ and $c_s^2$ are positive during the
evolution. The universe will be matter-dominant in the future.}
\label{lmzaandcs2}
\end{center}
\end{figure}

We performed the numerical calculations of the background evolution
and the coefficients for perturbations for different parameter choices in Figs.
\ref{lmzrhoandw} and \ref{lmzaandcs2}. In both figures, we have used the unit $8\pi G=1$
and chose the parameter $c_1=3.0\times 10^{122}$. The choices of parameter $c_2$ are different for these two cases.
In the first figure $c_2=0$ and in the second one $c_2=0.05$. In both cases we set the initial conditions as $\phi_i=-1.2$ and $\dot\phi_i=10^{-61}$
well within the matter dominated era in order to make the evolutions of dark energy consistent with the observations.
We can see that in both cases
the equation of state of dark energy crosses the cosmological
constant boundary, and the present values of
$w_{DE}$ and $\Omega_{DE}$ fits well with the observational
data. At present time $\ln a=0$, $w_{DE}=-1.158$ and $-1.167$ for the first and second cases respectively. These are consistent
with the current constraint by the observational data from CMB, LSS and Supernovae. For example the analysis in Ref. \cite{Liu:2010ba}
gives $w_{DE}=-1.143\pm 0.160$.
The values of $\Omega_{DE}$ in both cases are around $0.725$, also compatible with the observations \cite{Komatsu:2010fb}.
Moreover, the parameter $A$ defined in
(\ref{a}) and the sound speed squared defined in (\ref{cs2}) are
both positive during evolutions in these two cases. This means that
at the crossing there is no ghost instability and the perturbation
is regular and classically stable.

Finally we should emphasize that
with different parameters, these two cases give rise
to predictions of different fate of dark energy in the future. For
the first case where $c_2=0$ which reduces to the case discussed in
\cite{Deffayet:2010qz}, $w_{DE}$ crosses $-1$ and approaches $-2$ at early time, then when the dark energy dominates the universe
$w_{DE}$ will approach to $-1$ forever and the universe enters into the De Sitter phase.
This can be understood as follows. With vanished $c_2$, the model is symmetric under the field shift $\phi\rightarrow \phi+C$ with constant $C$.
The corresponding Noether current is
\be
J^{\mu}=c_1\nabla^{\mu}X+\nabla^{\mu}\phi-c_1\Box\phi\nabla^{\mu}\phi~,
\ee
and the conservation law $\nabla_{\mu}J^{\mu}=0$ is nothing but the equation of motion. For the background,
$\phi$ is homogeneous the conservation law simplifies as
\be
\dot J^0+3HJ^0=0~,
\ee
and the ``charge density" $J^0=\dot\phi-3c_1H\dot\phi^2$, which scales as $a^{-3}$. From it we can solve for $\dot\phi$
that
\be\label{dotphi}
\dot\phi=\frac{1\pm \sqrt{1-12c_1H J^0}}{6c_1H}~.
\ee
On the other hand the energy density and pressure are respectively
\bea
\rho&=&(-1+6c_1H\dot\phi)\frac{\dot\phi^2}{2}\nonumber\\
p&=&-(1+2c_1\ddot\phi)\frac{\dot\phi^2}{2}~.
\eea
To guarantee the positivity of the energy density, $6c_1H\dot\phi>1$, in Eq. (\ref{dotphi}), we should choose the plus sign, i.e.,
\be
\dot\phi=\frac{1+ \sqrt{1-12c_1H J^0}}{6c_1H}~.
\ee
Because $J^0$ scales as $a^{-3}$, soon after the beginning $J^0$ approaches to zero so that $\dot\phi\rightarrow \frac{1}{3c_1H}$ and
$\ddot\phi\rightarrow -\frac{\dot H}{3c_1 H^2}$. This time the universe is still dominated by matter and the Hubble rate $H$ scales as $a^{-3/2}$,
it is easy to obtain that $\ddot\phi$ approaches to the constant $\frac{1}{2c_1}$.
So the equation of state
\be
w_{DE}=-\frac{1+2c_1\ddot\phi}{-1+6c_1H\dot\phi}\rightarrow -(1+2c_1\ddot\phi)=-2~.
\ee
This is a phantom attractor solution to the model. At this period, the energy density of the dark energy increases as fast as $a^3$ and the matter
energy density decreases as $a^{-3}$. So with the initial conditions we have chosen the universe transits to the phase of dark energy domination at
low redshift. In this phase $\dot\phi$ is more close to $\frac{1}{3c_1H}$ than in the matter dominated era. However, the Hubble rate is
\be
H^2\simeq  {1\over 3}(-1+6c_1H\dot\phi)\frac{\dot\phi^2}{2}\rightarrow \frac{\dot\phi^2}{6}\rightarrow \frac{1}{54c_1^2 H^2}~,
\ee
so $H$ approaches to the constant $(54c_1^2)^{-1/4}$ and $\ddot \phi\rightarrow 0$. The equation of state $w_{DE}\rightarrow -(1+2c_1\ddot\phi)
\rightarrow -1$. The universe enters into the de Sitter phase in the future.

For the second case where $c_2=0.05$, the shift symmetry is violated by the ``effective" mass term.
The ``charge density" satisfies the equation
\be
\dot J^0+3HJ^0=-c_2\dot\phi^2\phi~,
\ee
where the density $J^0$ gets a minor modification compared with the first case,
\be
J^0=(1-c_2\phi^2)\dot\phi-3c_1H\dot\phi^2~.
\ee
With the parameters and the initial conditions we have chosen, the effective mass term is very small at the time of matter domination.
The early evolution of dark energy in this case is similar to the first case, its equation of state crosses $-1$ and approaches to the
phantom attractor solution $w_{DE}=-2$. Because the initial value of $\phi$ we have chosen is negative and $\dot\phi\simeq \frac{1}{3c_1 H}>0$,
$\phi$ increases from the negative region to the positive one, so the effective mass term remains small during a fairly long period.
Recently when the universe shifts to the dark energy dominated phase, $w_{DE}$ increases from $-2$ and approaches $-1$. However, the value of the field
$\phi$ increases continually and the effective mass term becomes more and more important. So the universe will not enter into the de Sitter phase and
$w_{DE}$ increases continually and gets back to larger than zero again.
When $c_2\phi^2$ increases to the value much larger than unity, it will dominate the energy density and the pressure of the dark energy as seen from
Eq. (\ref{densityp}). The equation of state $w_{DE}\rightarrow 1$, but this phase is not stable because the energy density of the
dark energy will decrease as fast as $a^{-6}$.
Hence in the future dark energy might exit
domination and the universe will return to matter-dominant stage.
This phenomenon of the model has not been discussed before.


\section{Summary}

The model building of dark energy which is possible to cross the cosmological constant boundary, i.e., the quintom dark energy
has attracted a lot of attention in the literature
(for example see Refs. \cite{Quintom_sum}). A very interesting question is how to build a single scalar model
which gives rise to the crossing without any instabilities. The past studies have shown that a single scalar field
satisfying a generic k-essence Lagrangian cannot give rise to the stable violation of NEC. However, it becomes
possible when higher derivative operators are introduced. The non-degenerate higher derivative
model can eliminate the classical instabilities but is still plagued by the ghost mode.
In this paper we considered the degenerate higher derivative model inspired by
the Galileon theory and its generalizations. We
started with a single scalar of which the Lagrangian contains generic higher derivative operator,
and then suggested a degenerate condition to eliminate the extra degrees of freedom brought by these higher derivative
operators. Using the degenerate condition in the curved spacetime, the Lagrangian can be reduced to the form (\ref{kg}). We studied the cosmological
application of this model, including the background and perturbation analysis. Particularly, we chose an explicit example
and performed a detailed numerical computation. Our numerical results verified that this model is able to realize the EoS
to cross the cosmological constant boundary and the perturbation evolves smoothly without any pathologies.

A quintom model which successfully violates the NEC without leading
to quantum instability also has important implications to the early
universe physics. In Refs. \cite{Cai:2007}, the authors found that a
quintom model can avoid the big bang singularity widely existing in
standard and inflationary cosmologies. In this picture, the moment
of initial singularity can be replaced by a big bounce. Based on
this scenario, the corresponding perturbation theory has been
extensively developed, namely, on adiabatic perturbations
\cite{Cai:2008qb, Cai:2008qw, Cai:2009hc},
non-Gaussianities\cite{Cai:2009fn}, entropy
fluctuations\cite{Cai:2008qw, Cai:2011zx}, and the related
preheating phase\cite{Cai:2011ci}. Recently, it was observed that
the Galileon model with the EoS across $-1$ exactly leads to a
bouncing solution in the frame of the flat FRW
universe\cite{Qiu:2011cy}. Consequently, we expect that bouncing
cosmologies can be realized in a generic quintom model with
degenerate higher derivative operators.

\section{Acknowledgement}

The research of ML is supported in part by National Science
Foundation of China under Grants No. 11075074 and No. 11065004, by
the Specialized Research Fund for the Doctoral Program of Higher
Education (SRFDP) under Grant No. 20090091120054 and by SRF for
ROCS, SEM. TQ is supported by Taiwan National Science Council (NSC)
under Project No. NSC98-2811-M-002-501 and No. NSC98-2119-M-002-001.
YC is supported by funds of physics department at Arizona State
University. XZ is supported in part by the National Science
Foundation of China under Grants No. 10821063, 10975142 and
11033005, and by the Chinese Academy of Sciences under Grant No.
KJCX3-SYW-N2.

{}

\end{document}